\documentclass[14pt]{article}

\usepackage{arxiv}

\usepackage[utf8]{inputenc} 
\usepackage[T1]{fontenc}    
\usepackage{url}            
\usepackage{booktabs}       
\usepackage{amsfonts}       
\usepackage{nicefrac}       
\usepackage{microtype}      
\usepackage{lipsum}
\usepackage{amssymb,amsmath,amsfonts,bbm,eurosym,ulem,graphicx,caption,color,setspace,comment,footmisc,caption,pdflscape,subfigure,array}
\usepackage{algorithm, algorithmicx}
\usepackage{algpseudocode}
\usepackage{float}
\usepackage{appendix}
\usepackage{xcolor}
\usepackage[colorlinks = true,
            linkcolor = blue,
            urlcolor  = blue,
            citecolor = blue,
            anchorcolor = blue]{hyperref}

\usepackage{graphicx}
\graphicspath{ {./images/} }

\newtheorem{theorem}{Theorem}

\newcolumntype{L}[1]{>{\raggedright\let\newline\\arraybackslash\hspace{0pt}}m{#1}}
\newcolumntype{C}[1]{>{\centering\let\newline\\arraybackslash\hspace{0pt}}m{#1}}
\newcolumntype{R}[1]{>{\raggedleft\let\newline\\arraybackslash\hspace{0pt}}m{#1}}
\DeclareMathOperator*{\argmin}{argmin}

\title{Bayes Optimal Informer Sets for Early-Stage Drug Discovery}

\author{
Peng Yu\\
    Department of Statistics\\
    University of Wisconsin-Madison\\
    Madison, WI, 53706\\
    \texttt{peng.yu@wisc.edu}\\
\And
Spencer S. Ericksen\\
    UW-Carbone Cancer center\\
    School of Medicine and Public Health\\
    University of Wisconsin-Madison\\
    Madison, WI, 53706\\
\And
Anthony Gitter\\
    Department of Biostatistics and Medical Informatics\\
    University of Wisconsin-Madison\\
    Morgridge Institute of Research\\
    Madison, WI, 53706\\
\And
Michael A. Newton\\
    Department of Statistics\\
    Department of Biostatistics and Medical Informatics\\
    University of Wisconsin-Madison\\
    Madison, WI, 53706\\
    \texttt{newton@biostat.wisc.edu}\\
}

\begin{document}
\maketitle
\begin{abstract}
 An important experimental design problem 
 in early-stage drug discovery is 
how to prioritize available compounds for testing when very little is known about the
target protein. Informer based ranking (IBR) methods address the prioritization problem
when the compounds have provided bioactivity
data on other potentially relevant 
targets.  An IBR method
selects an informer set of compounds, and then prioritizes the remaining compounds 
on the basis of new bioactivity experiments performed with the informer set on the target. 
 We  formalize the problem as a two-stage decision problem and introduce the Bayes Optimal Informer SEt (BOISE) method for its solution. 
 BOISE leverages a flexible model of the initial bioactivity data, a relevant loss function,
 and effective computational schemes
  to  resolve the two-step design problem.  We evaluate BOISE and compare it to other IBR
  strategies in two retrospective studies, one on protein-kinase inhibition and 
  the other on anti-cancer drug sensitivity. In both empirical settings BOISE exhibits 
  better predictive performance than available methods.   It also behaves well with missing data, where
   methods that use matrix completion show
  worse predictive performance.  We provide an R implementation of BOISE at \url{github.com/wiscstatman/esdd/BOISE}.
\vspace{0.1in}
\end{abstract}





\section{Introduction} \label{sec:introduction}

Chemical screening laboratories, such as core service facilities in academic medical centers, are often faced with the following experimental design problem.  
Having received some quantity of a purified protein target, they must plan and deploy experimental assays to identify which available compounds produce some desired effect (bioactivity) on the protein's function.
Na\"ive high-throughput screening (HTS), a ``brute-force" approach in which massive, fixed collections of drug-like compounds are tested exhaustively, is often too expensive and risky for academic investigators.
Less expensive  alternatives to na\"ive HTS are needed for early-stage drug-discovery efforts, as in such cases where the risk merits limited resource commitment, e.g., a target protein's therapeutic relevance has not been fully validated.
Computational strategies that effectively prioritize compounds can reduce the amount of experimentation  required to find active compounds. 
Compared to HTS, such strategies  can also accommodate rapidly expanding accessible chemical space, where billions of virtual compounds are now readily synthesized and purchased on-demand.
Thus, bioactivity experiments are frequently preceded by virtual screening calculations that utilize information in novel ways, such as through molecular docking computations when protein structures are available (e.g., Souza et al. 2020), or through machine learning computations that build predictive models based on compound testing data (e.g., Sliwoski et al. 2014; Liu et al. 2018; Bajorath et al. 2020). 
However, these established approaches depend on structural or bioactivity data that are typically limited for novel targets. 

Within this large research domain, we focus on an extreme though not uncommon case involving a novel target from a well-characterized class of targets.
Here very little information is available on the target protein, beyond the knowledge that the target is from a class of proteins for which bioactivity has been measured on some common set of molecules. 
We address the specific challenge problem described in Zhang {\em et al.} (2019) to deploy {\em informer based ranking} (IBR).  
IBR selects a small subset of drug-like compounds (informers) from the common set that, upon testing against a new target, provides sufficient information to enable bioactivity predictions for the remaining untested compounds.
The predictions are used to rank-order the remaining compounds for testing (Figure 1)--rather than exhaustively testing the set.
\begin{figure}[!ht]
\centering
\includegraphics[width=6.0in]{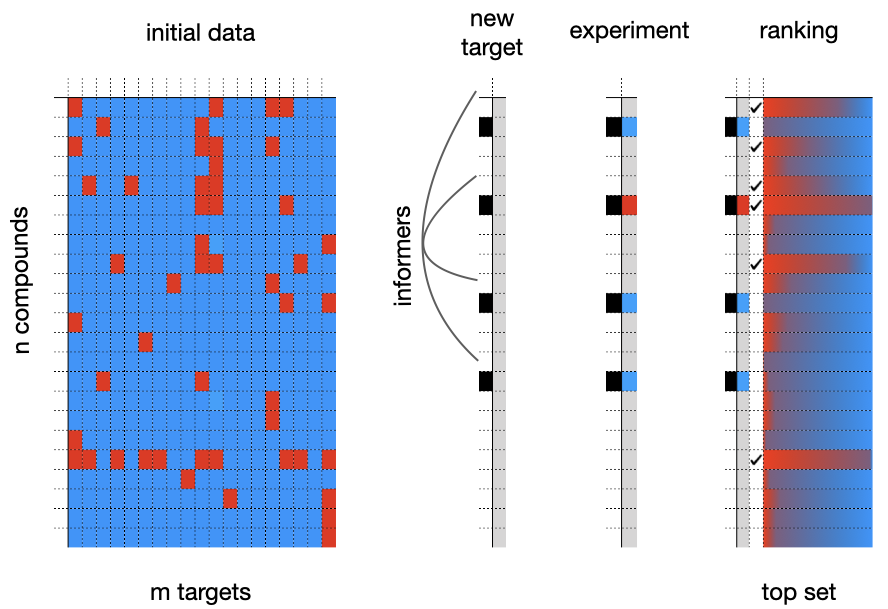}
\caption{\label{fig:scheme} 
{\bf Informer-based-ranking problem:} A matrix of binary bioactivity data 
is available (left; red active, blue inactive).  The problem is to first identify a subset of the $n$  compounds as informer compounds
that will be evaluated experimentally on new target, and then to prioritize
all the compounds for further testing after intermediate
data is obtained. }
\end{figure}
It is illuminating to recognize that an IBR 
strategy attempts to solve a two-stage finite statistical decision problem (e.g.,
Wald, 1950; Parmigiani and Inoue, 2009, page 230).
There is first the question of how to constitute the small set 
of informer drugs and then the question of what to do with intermediate data
measured on these drugs in order to prioritize the 
remaining compounds. In the present work
we compare available IBR strategies to a novel strategy developed from the decision-theoretic perspective.  

 In Zhang {\em et al.} (2019), 
 domain-specific baseline IBR's were compared to alternatives guided by heuristics and machine learning.
 Baseline methods include informer-set selection by the frequent-hitters rule, which selects compounds showing activity against the most targets in the initial data set.
 Baseline chemometric strategies, by contrast, use available distances computed between compounds in chemical space to choose a chemically diverse informer set. 
The machine-learning strategies partition the bioactivity data, producing  clusters of relatively similar targets;  then informer compounds are  selected as those  predictive of  cluster label.  
Since these effective IBR strategies leverage statistical
patterns in the  bioactivities, we reason that
statistical modeling may provide a useful approach to deriving
more effective  strategies than are currently available. 

To develop an IBR method using decision theory,  
consider a thought-experiment suggested by sequential
analysis. 
If in addition to the initial bioactivity data we knew the identity of the informers as 
well as their bioactivity measurements against the new target,
then we 
would be well positioned to rank the as-yet-untested compounds, say by their posterior expected activity in the context of a statistical model. 
But we know neither the informers nor the intermediate data
they would provide.  
With a model we could consider the predictive distribution of intermediate data on any candidate informer set; indeed we could imagine simulating this predictive distribution given the initial bioactivities.  
In each simulated instance we would have sufficient information to rank the as-yet-untested compounds, and 
by some form of averaging we could assess the expected loss tied to this candidate set.  
By similarly scoring any candidate set we would obtain an objective function whose optimization provides the best possible informers in the context of the chosen sampling model and loss function.  
Such Bayes Optimal Informer SEts (BOISE)  are candidate compound sets that minimize an average loss computed on hypothetical intermediate data. 
To produce an effective and practical IBR scheme we need a flexible sampling model, a discriminating loss function, and a nimble algorithmic approach, which we propose in Section~2.  

We evaluate BOISE retrospectively using  
the human protein kinase data set PKIS1, (Drewry {\em et al.} 2014), and  the anti-cancer drug sensitivity data set GDSC1 (Yang {\em et al.} 2013).
BOISE performs better than other IBR schemes in predicting compound activity from both complete and incomplete initial bioactivity data.


\section{Methodology}

\subsection{Problem Setting}

We are given an $m \times n$  matrix, denoted $x_0$, that contains bioactivity data measured on $m$ targets and $n$ compounds, and we  denote the initial targets as $I = \{1,2,\cdots,m\}$, and the set of available compounds as $J=\{1,2,\cdots,n\}$.  Our studies have  considered  data sets with $m$ and $n$ in the hundreds, though larger systems are quite relevant. Suppressing the `0' subscript, 
we use  $x_{i,j}$ to 
denote the $i/j$'th entry of matrix $x_0$.  Thus $x_{i,j}$ is the outcome of a bioactivity
experiment involving protein target $i$ and drug compound $j$.  We treat the simplest case in the
present paper, taking binary data:
$x_{i,j} = 1$ indicates that compound $j$ is inferred to
be active on target $i$ while $x_{i,j} = 0$ 
corresponds to inactivity.  Much of the subsequent development 
is also relevant to quantitative bioactivity data.


Key to the problem is a new protein target, labeled
 $i^* \notin I$, on which   we have no bioactivity data at the outset of the experiment.   
We seek a relatively small set of compounds, $A \subset J$, the
choice of which will be guided by $x_0$.  Experimentation 
will be performed to assess the bioactivity of compounds in this informer set $A$ against target $i^*$, resulting in intermediate data 
$x_{A} = \left\{ x_{i^*,j}: j \in A \right\}$. 
Taken
together, $x_0$ and $x_A$
are  used to prioritize
other compounds for further testing.
One way to formalize this step is to  suppose that we must select a final top set 
$T = T\left( x_0, A, x_A \right) \subset J$
on which we will perform further experimentation  in order to identify as many compounds as possible with bioactivity against the target $i^*$.  We are
thinking of scenarios where we cannot screen the 
entire set $I$ (otherwise there's no need for
an informer set).
Design parameters here include the cardinality of the informer set, say $n_A$, and the cardinality of the top set, $n_T$.  

Our statistical analysis rests on an elementary sampling model, namely that 
bioactivity data $\{ x_{i,j} \}$ are realizations of 
mutually independent Bernoulli trials when conditioned on corresponding parameters
$\theta=\{ \theta_{i,j} \}$.  Further modeling will constrain these parameter values so that
information may be readily shared among compounds and targets, but the Bernoulli observation
component anchors the entire approach. We think of each
$\theta_{i,j} = P(x_{i,j}=1|\theta_{i,j})$ as a true bioactivity level balancing biological and technical
variation of assays that measure the effect of compound $j$ on protein $i$. Roughly speaking, we seek compounds $j$ for
which $\theta_{i^*,j}$ is large, and this goal is conveniently encoded by the proposed loss
function:
\begin{eqnarray}
\label{eq:loss}
L(A, T, \theta) = \sum_{j \in T} \left( 1- \theta_{i^*,j} \right).
\end{eqnarray}
Because inference requires both an informer set $A$ and a top set $T$,
we express the loss in terms of these two actions as well as the full state of nature
$\theta$.  The loss function would be trivial to minimize if parameters $\theta$
were known, but barring this
we elaborate the model and pursue actions to minimize an appropriate average loss.

\subsection{Bayes optimal IBR}

We are  guided by Bayesian statistical decision theory (Berger, 1985;
Robert, 2007; Parmigiani and Inoue, 2009). Relative to a to-be-specified
prior distribution $p(\theta)$, the Bayes risk of the two-stage rule
$\{A, T\}=\{A(x_0),T(x_0,A,x_A)\}$ is 
the marginal expected loss, averaging~(\ref{eq:loss}) over the multi-Bernoulli
sampling model $p(x_0|\theta)$ and $p(x_A|\theta)$ as well as over the
prior $p(\theta)$.
Using integral notation for sums over respective sample
or parameter spaces, and considering $A$ and $T$
as functions on their input, the (marginal) Bayes risk is
\begin{eqnarray}
\label{eq:risk}
r(A, T) 
= \int \int \int  L\left\{ A(x_0), T(A,x_A,x_0), \theta \right\} \,
 p(x_0| \theta) \, p(x_A| \theta) \, p(\theta) \, d\theta \, d x_A \, d x_0.
\end{eqnarray}
 An inference procedure that
minimizes the Bayes risk is called a Bayes rule; 
in the context of the model, the prior and loss, its use is
a rational way to design and carry out the experiment.

Our first finding concerns a simplification of the Bayes risk for the particular
loss function~(\ref{eq:loss}).  We place ourselves at the point at which 
we have named informer compounds
$A$ and have received intermediate bioactivity data $x_A$; with
modeling components at hand, we could compute the posterior mean
$\hat \theta_{i^*,j} = \mathbb{E}\left( \theta_{i^*,j} |x_A, x_0 \right)$
by averaging in the posterior distribution
\begin{eqnarray}
\label{eq:doublepost}
p(\theta|x_0,x_A) =  p(x_0|\theta) \, p(x_A| \theta) \, p(\theta)/ p(x_0, x_A) .
\end{eqnarray}
We define  top set rule $T^*$ to select $n_T$ compounds
having  largest posterior means $\hat \theta_{i^*,j}$, with ties broken arbitrarily
if necessary.  This set turns out to be the Bayes rule for the sub-problem to identify a top-set
(e.g., Henderson, 2015, page 17), which we use to confirm:
\begin{theorem}
For any rules $A$ and $T$, $r(A,T) \geq r(A, T^*)$.
\end{theorem}
The lower bound above is the Bayes risk associated with the best possible top-set rule for any
given informer-set $A$.  Essentially, this shows how to score any informer-set rule by
profiling out the top-set selection. 
The risk~(\ref{eq:risk}) is amenable to further simplification by formally integrating
the parameters $\theta$:
\begin{eqnarray*}
r(A,T) \geq r(A,T^*) 
 = \int_{x_0} {\mbox {\rm PEL}}_1 \left(x_0, A \right) \, p(x_0) \, d x_0.
\end{eqnarray*}
PEL stands for posterior expected loss, and the subscript is meant to indicate that the
distribution is posterior to the initial activity data $x_0$:
\begin{eqnarray}
\label{eq:pel}
{\mbox {\rm PEL}}_1(x_0,A) = \int_{x_{ A}} p\left( x_{A} | x_0 \right)
\left\{\sum_{j \in  T^*(x_0,A,x_A)} ( 1 - \hat \theta_{i^*,j} ) \right\} \, dx_{ A} 
\end{eqnarray}
Using a standard result from  decision theory, the rule $A$ 
that minimizes the marginal Bayes risk is obtained by
finding the best informer set at each $x_0$ (e.g., Berger, 1985, page 159).  Thus,
the Bayes optimal informer set  $A^*(x_0)$  is
\begin{eqnarray}
\label{eq:boise}
A^*(x_0) = \argmin\limits_{A \subset J, \, |A| = n_A}  \, {\mbox {\rm PEL}}_1(x_0,A).
\end{eqnarray}
It is useful to name the quantity in braces
 in~(\ref{eq:pel}), for this too is a posterior expected loss, but 
conditional on both $x_0$ and $x_A$, and utilizing the top-set rule $T^*$.  
We denote it by PEL$_2(x_0,A,x_A)$ and note:
\begin{eqnarray}
\label{eq:pel2}
{\rm PEL}_1(x_0,A) = \mathbb E\left\{ {\mbox {\rm PEL}}_2(x_0,A, x_A) \, |\, x_0\right\},
\quad
{\mbox {\rm PEL}}_2(x_0,A,x_A) =  \sum_{j \in  T^*(x_0,A,x_A)} ( 1 - \hat \theta_{i^*,j} ).
\end{eqnarray}
With these facts a general program is beginning to emerge (Algorithm~\ref{Algo0}).
We may score  any candidate informer set $A$ by
PEL$_1$, which is computed as an average of PEL$_2$ scores, possibly obtained by
sampling the predictive distribution of intermediate data $x_A$.  In other words, we stochastically
predict what intermediate data would emerge if we were to use informer set $A$, and we average the
further expected loss associated with optimal top-set construction from those completed data.  By varying $A$
we find the Bayes Optimal Informer SEt (BOISE) associated with the least average loss.  Our 
logic parallels dynamic programming for sequential decision analysis, from early 
developments in multi-stage finite decision problems (Wald, 1950) to more recent work in clinical trials (Berry, 2006).
\begin{algorithm}
\caption{Compute posterior expected loss of a candidate informer set \label{Algo0}}
\hspace*{\algorithmicindent} \textbf{Input:} Initial data $x_0$, candidate informer set $A$, size of top set $n_T$, model structure\\
\hspace*{\algorithmicindent} \textbf{Output:} Monte Carlo approximation to posterior expected loss PEL$_1(x_0,A)$
\begin{algorithmic}[1]
\While{predictive sampling}
\State sample $\theta$ from $p(\theta|x_0)$
\State sample $x_A$ from $p(x_A| \theta)$
\State for all compounds $j$ compute future posterior summary $\hat \theta_{i^*,j} = E\left(\theta_{i^*,j}|x_0,x_A \right)$
\State form  top set $T^*(x_0,A,x_A)$ holding $n_T$ compounds with largest $\hat 
 \theta_{i^*,j}$
\State compute PEL$_2(x_0,A,x_A) = \sum_{j \in T^*(x_0,A,x_A)} (1 - \hat \theta_{i^*,j} )$
\EndWhile
\State average PEL$_2$ scores to have approximate PEL$_1$
\end{algorithmic}
\end{algorithm}

\subsection{Modeling the parameter space}

Specific BOISE schemes depend on the
configuration of probability over
parameters $\theta = \{ \theta_{i,j} \}$.   Allowing too much flexibility 
 limits the utility of initial data $x_0$ to predict  anything about the new target
$i^*$.  On the other hand, an overly restrictive model is liable to miss important bioactivity
signatures. Also, a model supporting feasible  computations 
is especially critical for contemporary applications.
We pursue a theme proposed in Zhang {\em et al.} (2019) to
cluster the target space, and for our primary calculations we develop this theme using  techniques from nonparametric Bayesian
analysis  (e.g., Hjort {\em et al.}, 2010).  

To retain flexibility
while controlling the parameter-space complexity, we assume there is 
a partition $\mathcal{C} = \{ c_k \}$ of the $m$ initial targets, wherein each cluster $c_k$ contains 
identically distributed targets in the sense that
$\theta_{i,j} = \phi_{k,j}\mathbbm{1}(i\in c_k)$
for a reduced set of cluster/compound parameters $\phi=\{ \phi_{k,j} \}$.   Furthermore, 
we propose three positive hyper-parameters $m_0, \alpha_0,$ 
and $\beta_0$ to control the probability
distribution over $\phi$ and $\mathcal{C}$, which: (1) encodes independence between cluster structure and activity rates, 
(2) has all entries of $\phi$ being mutually independent Beta$(\alpha_0,\beta_0)$, and (3)  governs
partition $\mathcal{C}$ by a Chinese-Restaurant distribution:
\begin{eqnarray}
\label{eq:crm}
p(\mathcal{C}) = \frac{ m_0^K \Gamma(m_0) }{ \Gamma(m+m_0) } \prod_{k=1}^K
\Gamma( m_k ) .
\end{eqnarray}
Here $m_k$ counts the number of targets in $c_k$ and  $\mathcal{C}$ 
is composed of $K$ clusters.  
We say the distribution encoded in~(\ref{eq:crm}) is CR$_m(m_0)$. The proposed  specification 
for the initial bioactivity data $x_0$ is thus:
\begin{align}
\label{eq:modelstructure}
\begin{split}
\mathcal{C} &\sim {\rm CR}_m(m_0), \\  
    \phi_{k,j} & \sim {\rm Beta}(\alpha_0,\beta_0), \quad  k=1,\cdots, K, \; j=1,\cdots,n \\
x_{i,j} \mid  \mathcal{C}, \phi  & \sim {\mbox {\rm Bernoulli}} \left\{ \phi_{k,j} \mathbbm{1}(i\in c_k) \right\},
 \quad i=1,\cdots,m, \; j=1,\cdots,n.
\end{split}
\end{align}
An exchangeable connection to bioactivities $x_{i^*,j}$ on the new target $i^*$ is available immediately.
 CR$_{m+1}(m_0)$ would assert that given $\mathcal{C}$, the new target $i^*$ becomes 
part of cluster $k$ with probability proportional to $m_k$, in 
which case we write
$i^* \rightarrow c_k$.  It 
populates a cluster by itself with probability proportional to $m_0$.  If $i^* \rightarrow c_k$, then $x_{i^*,j}$ is Bernoulli$( \phi_{k,j} )$ like the other
targets in that cluster.  If $i^*$ populates a new cluster, say $c_0$, then there must be some other
rates $\phi_{0,j}$ governing these Bernoulli trials, and these rates themselves are distributed by
the same Beta$(\alpha_0,\beta_0)$ distribution.

Model~(\ref{eq:modelstructure}) is quite flexible, allowing
that target $i^*$ has a bioactivity pattern in common
with some subset 
of initial proteins, and accounting for uncertainty in this 
cluster subset. 
There could be further benefit to
clustering in the compound space or to adopting a more
elaborate specification, though 
the direction taken is suggested by the
inference task, which focuses on new targets for the fixed
set of compounds. Also, computations
appear to be considerably more difficult in elaborations
of the present case.  One advantage of~(\ref{eq:modelstructure}) is that 
explicit integration eliminates all the $\phi_{j,k}$ parameters, much like 
as happens for  collapsed Gibbs sampler computations in a related context (Liu, 1994).  Therefore,
the predictive sampling in Algorithm~\ref{Algo0} entails the sampling of clusterings
$\mathcal{C}$ rather than fully elaborated parameter states $\theta$.  A second advantage of~(\ref{eq:modelstructure}) comes from how it meshes with the loss function~(\ref{eq:loss}).  
PEL$_2$ (and thus PEL$_1$) calculations generally require averaging with respect to the posterior distribution
$p(\theta|x_0,x_A)$ as in~(\ref{eq:doublepost}), which in more elaborate specifications may require 
posterior sampling under each simulated $x_A$.   For model~(\ref{eq:modelstructure}) we find a scheme to obtain PEL$_1$ 
via posterior sampling from $p(\mathcal{C}|x_0)$  and 
predictive sampling of $p(x_A|x_0)$, but in which no sampling conditional upon $x_A$ is required.

\subsection{Computations}

In the context of model~(\ref{eq:modelstructure}), the general program (Algorithm~\ref{Algo0}) becomes more explicit.  Sampling of intermediate bioactivity states $x_A$ may be arranged by first sampling clusterings $\mathcal C$ from 
$p(\mathcal C|x_0)$ and then drawing data from $p(x_A|x_0,\mathcal{C})$, recognizing the simplified form
\begin{eqnarray}
\label{eq:intermediate}
 x_{i^*,j}\,|  \mathcal{C}, x_0 \, \sim_{\rm ind}\,{\rm Bernoulli} \left(\frac{a_{k,j}}{a_{k,j}+b_{k,j}}\right) \qquad 
  {\rm when} \;  i^* \rightarrow c_k,
\end{eqnarray}
where $a_{k,j} = \alpha_0 + \sum_{i \in c_k}x_{i,j}$ and $b_{k,j} = \beta_0 + \sum_{i\in c_k}(1-x_{i,j})$ counts actives and inactives, respectively. 
From CR$_{m+1}(m_0)$, the probability that $i^* \rightarrow c_k$
is $m_k/(m+m_0)$, conditionally on $\mathcal{C}$ and $x_0$.  There is also probability $m_0/(m+m_0)$ 
that the new target  $i^*$ does not cluster with the initial targets, in which case the bioactivities
$x_{i^*,j}$ are i.i.d. Bernoulli$\left\{\alpha_0/(\alpha_0 + \beta_0)\right\}$.
It is convenient to represent the new singleton cluster as $c_0$, and set
$a_{0,j}=\alpha_0$ and $b_{0,j}=\beta_0$.

A difficult aspect of this predictive sampling scheme is how to draw clusterings from $p(\mathcal{C}|x_0)$, which is analogous to calculations required in Dirichlet process mixture models (DPMMs).  A great deal of progress has been made on this general problem, and we tap into 
these nonparametric Bayesian results to advance our calculations.  Appendix~B presents a Gibbs sampler
adapted to the present context from MacEachern  (1994) and
Neal (2000).

The next computational challenge  is the 
evaluation of the optimal top
set $T^*(x_0,A,x_A)$, which holds the $n_T$ compounds having
the highest values of $\hat \theta_{i^*,j}=E( \theta_{i^*,j} | x_0, x_A )$.   Our approach is to
re-use the sampled clusterings $\mathcal{C}$, noting by iterated expectations
that 
$\hat \theta_{i^*,j} = E\left\{  \tilde \theta_{i^*,j} | x_0, x_A \right\},$  where 
$\tilde \theta_{i^*,j} = E( \theta_{i^*,j} | \mathcal{C}, x_0, x_A )$.  This inner expectation  is an average over ways the new target $i^*$ may (or may not) cluster with the existing targets,
and we find:
\begin{eqnarray}
\label{eq:thetatilde}
\tilde \theta_{i^*,j}  = \sum_{k=0}^K p_k \left\{ \frac{a_{k,j}+x_{i^*,j}\mathbbm{1}\left(j\in A\right)}{a_{k,j}+b_{k,j}+\mathbbm{1}\left(j\in A\right)} \right\}
\end{eqnarray}
where $p_k$ is the conditional probability (given $x_0$, $x_A$, and $\mathcal{C}$)
that $i^*$ links to cluster $c_k$:
\begin{eqnarray*}
p_k \propto m_k\prod_{j\in A} \left(\frac{a_{k,j}}{a_{k,j}+b_{k,j}}\right)^{x_{i^*,j}}\left(\frac{b_{k,j}}{a_{k,j}+b_{k,j}}\right)^{1-x_{i^*,j}}
\end{eqnarray*}
and where proportionality is resolved by $\sum_{k=0}^K p_k=1$.   The sought-after
$\hat \theta_{i^*,j}$ is marginal to uncertainty in clusterings but conditional on
intermediate data $x_A$.   The generic solution would be to re-apply MCMC sampling over 
clusterings for all the different posterior distributions, but we propose to recycle the sampled clusterings already available from $p(\mathcal C |x_0)$ through an importance-sampling argument.   These modeling and computational elements allow for refinement of  Algorithm~\ref{Algo0}, which we report as
Algorithm~\ref{Alg:pel} in Appendix~C.

Our final job is to find the optimal informer set $A^*$ in  problem~(\ref{eq:boise}),
which optimizes over  discrete, size-$n_A$ subsets of compounds $J=\{1,2, \cdots, n\}$. 
The complexity of  ${\rm PEL}_1(x_0,A)$ in~(\ref{eq:pel}) makes this challenging,
but there is  an effective greedy  method based upon adding one compound at 
a time to a sequentially growing solution~(Algorithm~\ref{Alg:greedy}). We provide an R implementation of the complete BOISE procedure at \url{github.com/wiscstatman/esdd/BOISE}.


\begin{algorithm}
\caption{Greedy Informer Selection}\label{Alg:greedy}
\hspace*{\algorithmicindent} \textbf{Input:} Initial data $x_0$, size of informer set $n_A$, size of top set $n_T$.\\
\hspace*{\algorithmicindent} \textbf{Output:} Selected informer set of length $n_A$
\begin{algorithmic}[1]
\State \textbf{Initialization}: Evaluate ${\rm PEL}_1(x_0,A)$ for all $|A|=1$. Let $A^* = \argmin_{|A|=1}{\rm PEL}_1(x_0,A) $.
\While{$|A^*| < n_A$}
    \State Evaluate ${\rm PEL}_1(x_0,A^*\cup\{j\})$ for each $j\in J\setminus A^*$.
    \State Let $A^* \gets A^*\cup\{j^*\}$, where $j^* = \argmin_{j}{\rm PEL}_1(x_0,A^*\cup\{j\})$.
\EndWhile
\State Return $A^*$ as selected informer set.
\end{algorithmic}
\end{algorithm}

\section{Empirical studies}
\subsection{Protein kinases}
Protein kinases comprise the second largest drug target class and are the primary target class for cancer therapeutics.
Protein kinases attach phosphate groups to regulatory sites on the surfaces of other proteins, thereby modulating their functions.
Discovering drugs that inhibit kinase activity is a problem of broad interest.  
To assess the operating characteristics of BOISE in this domain, we use a public kinase  data set, PKIS1 (Drewry {\em et al.} 2014), downloaded from  \href{https://github.com/SpencerEricksen/informers/blob/master/data/pkis1.csv}{CHEMBL}. 
After preprocessing, PKIS1 contains the bioactivity scores for $m=224$  kinase targets and $n=366$ drug compounds. 
The data are continuous measures of kinase inhibition; we threshold to binary active/inactive records using the 2-standard-deviation rule as applied in~Zhang {\em et al.} (2019); we also compare BOISE to baseline
and machine-learning methods reported in that work.
 
 A prospective evaluation would use the PKIS1 data as initial data $x_0$
 and identify an informer set $A$ of compounds to evaluate on a new target.  
 In place of this ideal study, we use cross validation in a retrospective
 design. We repeatedly drop out one kinase target, considering
 the retained proteins to provide data $x_0$,  and  the dropped-out target 
 to play the
 role of $i^*$,
 the novel target whose bioactivity data are initially hidden from the analyst.
 In each drop-out case, we apply BOISE to find informer compounds. 
 We use the available $x_{A}$ data on these informers as intermediate data that allows a prioritization of all compounds (taking advantage of documented experiments).   
 Formally, BOISE imagines that our job is to report a top set $T$ after processing intermediate data.  
 In addressing this it produces a ranking of all compounds according to $E( \theta_{i^*,j} | x_0, x_{A} )$, and this ranking can be evaluated, both by unveiling all the bioactivity data on $i^*$ to reveal measured active compounds, and by comparing to rankings produced by other IBR schemes. 
 
 Computational chemists use a variety of metrics to compare drug ranking methods, and we report two commonly used metrics to evaluate BOISE and other IBR schemes: (1) the area under the receiver operating characteristic curve (ROCAUC), and (2) the normalized enrichment factor at 10\% (NEF10). 
 After the fact, we label the compounds $j$ in order from the top of their IBR ranking; so in BOISE, $E(\theta_{i^*,1}|x_0, x_{A})$ is the largest posterior mean.   
 As we move down the ranked list, say with index $t=1,2\cdots n$, and in light of complete experimental data on the target $i^*$, we record the true positive rate TPR$(t) = \sum_{j=1}^t x_{i^*,j}/\sum_{j=1}^n x_{i^*,j}$ and the false positive rate FPR$(t)=\sum_{j=1}^t (1-x_{i^*,j})/\sum_{j=1}^n (1-x_{i^*,j})$. 
 The ROC curve plots TPR$(t)$ vs FPR$(t)$ as we vary the threshold $t$, and ROCAUC is the area under this curve. 
 Higher values, of course, correspond to prioritization schemes that put more of the truly active compounds near the top of the list.
 Alternatively, the normalized 10\% enrichment factor NEF10 is a metric that emphasizes behavior in the top 10\% of the ranking, and is a linear transformation of TPR$(\lfloor n/10 \rfloor)$ (see Appendix~D).

Figure \ref{fig:pkis} summarizes the predictive performance of BOISE and several published
IBR methods using two informer set sizes $n_A=8, 16$ and $n_T=36$. Note the ROCAUC uses the entire
ranking and NEF10 uses the top 10\%; 
there is no particular connection to the value $n_T$ used by BOISE.   We find quite low sensitivity of BOISE to the value of $n_T$ (data not shown). A formal statistical comparison  affirms what seems evident
from Figure~\ref{fig:pkis}, that BOISE has superior 
operating characteristics in this example.  Specifically, we fit a linear
model to each metric, including a factor for IBR method and a factor for 
protein target.  Figure~\ref{fig:CI_pkis} shows 95\% confidence intervals for
contrasts between BOISE  and other methods, adjusted for multiple pairwise comparison by Tukey's method (e.g., Bretz {\em et al.} 2010).

\begin{figure}[!ht]
\centering
\includegraphics[width=6.0in]{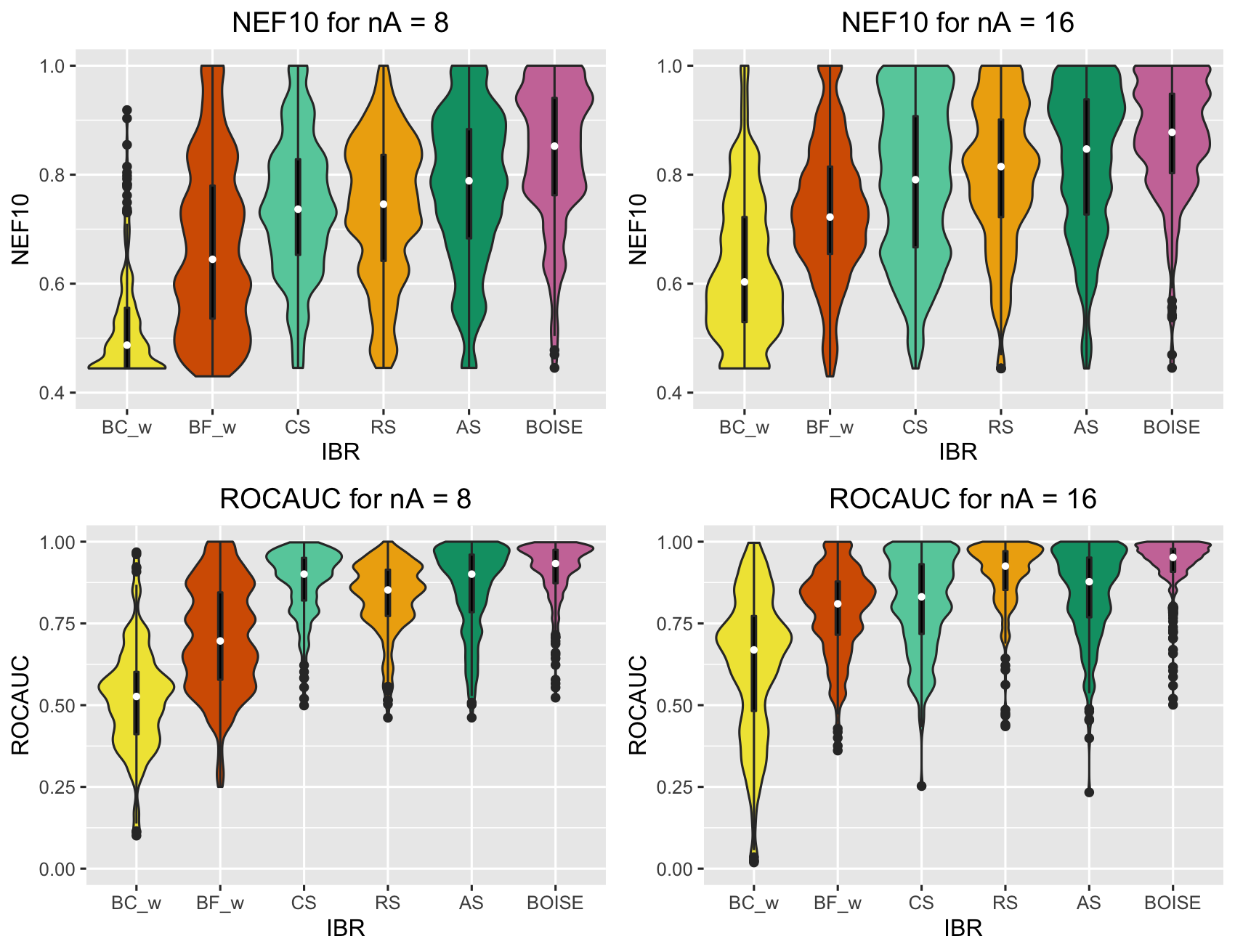}
\caption{\label{fig:pkis} 
{\bf Predictive performance of IBR methods on PKIS1 targets. } Two metrics (rows) are computed for each IBR method and on each leave-one-out
data set, and columns correspond to two choices
of informer-set size ($n_A=8, 16$). For both normalized enrichment (NEF10, top row) and ROCAUC (bottom row), a prioritization scheme that
puts all active compounds ahead of all inactive ones gets a perfect score (1.0); random guessing gets 0.5. 
 The median and interquartile ranges are displayed as a white circle and black bars, respectively, and violin
 plots show the empirical distribution of the metrics
 across the left-out targets.  IBR methods, distinguished by colors, are as described in the text: BC$_w$ and
 BF$_w$ are baseline schemes based on chemometric
 or frequent-hitters information; CS, RS, and AS are
 coding selection, regression selection, and
 adaptive selection, and BOISE is the proposed Bayes optimal 
 informer set method.}
\end{figure}

\begin{figure}[!ht]
\centering
\includegraphics[width=6.0in]{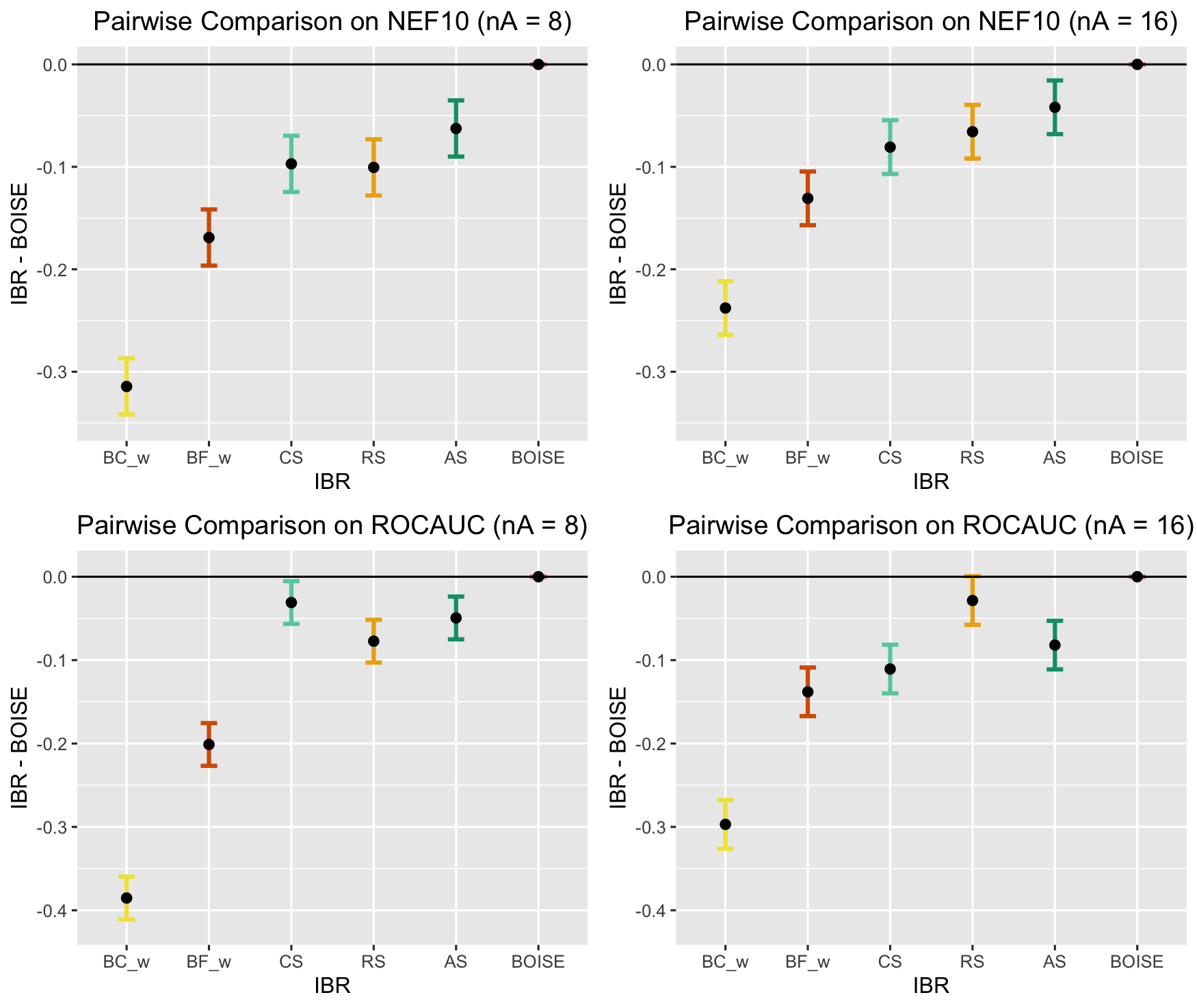}
\caption{\label{fig:CI_pkis} 
{\bf Statistical comparison of performance metrics on PKIS1 targets.}  
 Intervals are 95\% multiplicity adjusted confidence intervals of differences between 
 other IBR methods and BOISE, based on a linear model fit to the performance metrics 
 in Figure~2. }
\end{figure}

\subsection{Cancer cell lines}
We apply retrospective IBR calculations
using the Genomics of Drug Sensitivity in Cancer (GDSC) data set, downloaded from  \href{ftp://ftp.sanger.ac.uk/pub/project/cancerrxgene/releases/current_release/GDSC1_fitted_dose_response_25Feb20.xlsx}{Cancerrxgene} (Yang {\em et al.} 2013). It measures bioactivity of drugs
against cell lines derived from cancer tumors. A cell
line is obviously different than a purified protein, but
the experimental design problem is the same, to identify
an informer set of compounds that will be predictive of
other compounds' bioactivity against a new cancer.
GDSC reports standardized growth response data (z-scores) 
from  $304$ anti-cancer drugs and $987$ cancer cell lines.
We assume bioactivity, $x_{i,j}=1$, if the z-score is
less than -2.

\begin{figure}[!ht]
\centering
\includegraphics[width=6.0in]{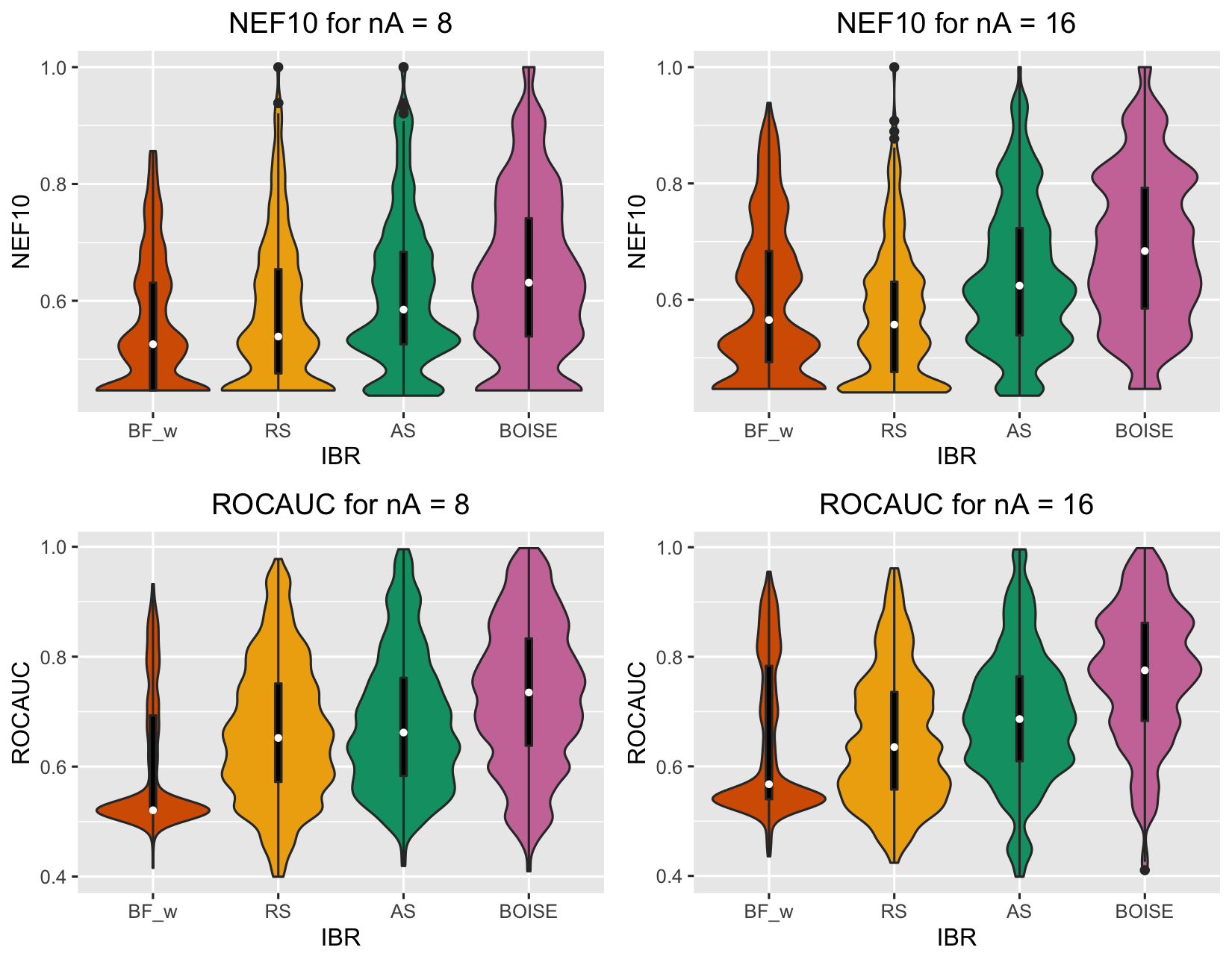}
\caption{\label{fig:gdsc} 
{\bf Predictive performance of IBR methods on complete subset of GDSC cell lines. }
  Metrics and details are as in Figure~2.  The complete GDSC subset
  includes 281 cell lines and 207 drug compounds; the empirical
  distributions of performance metrics are over the 281 left-out 
  cases. }
\end{figure}

\begin{figure}[!ht]
\centering
\includegraphics[width=6.0in]{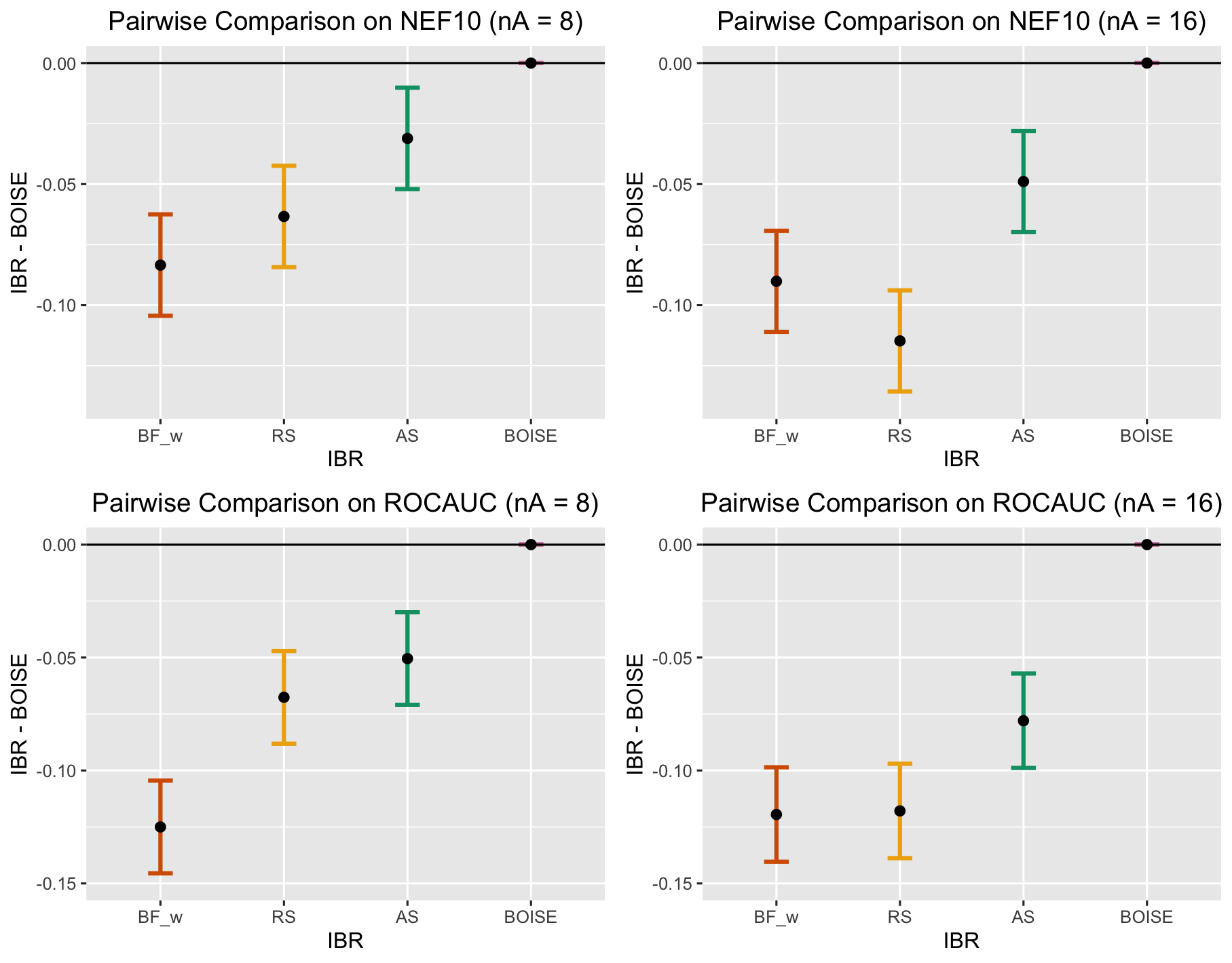}
\caption{\label{fig:CI_gdsc} 
{\bf Statistical comparison of performance metrics, GDSC.}  
 Intervals are 95\% multiplicity adjusted confidence intervals of differences between other IBR methods and BOISE, based on a linear model fit to the performance metrics 
 in Figure~4. }
\end{figure}

GDSC has a substantial amount of missing data (15.1\% of the matrix), which provides an opportunity to compare IBR
strategies in this context.  We report two numerical
experiments. In the first, we reduce GDSC to a
complete sub-matrix, which has 281 cell lines and 207 drugs found by removing rows/columns with more than $100$ missing entries.
We then do the same cross-validation exercise as we did with PKIS1 (though we drop the {\em coding selection} method 
due to its heavy computational cost).   Figure~\ref{fig:gdsc} 
shows that BOISE continues to have impressive metrics
of predictive performance, and this is confirmed in the confidence 
intervals in Figure~\ref{fig:CI_gdsc}.  For NEF10, for example, BOISE's method effect exceeds the closest competitor
by  $0.031$ units when $n_A = 8$ and $0.049$ when $n_A = 16$. 

Missing data in the initial matrix $x_0$ can present
a serious challenge to available IBR methods, which 
have to invoke some form of matrix completion as a pre-processing step. Missing data 
is not a fundamental problem for BOISE, considering
that probabilities calibrate whatever information is 
available. Two  aspects of BOISE  are influenced by missing data: cluster label updates in DPMM clustering, and posterior expectation calculation in informer selection. Both procedures rely on $a_{k,j}$ and $b_{k,j}$ in (\ref{eq:intermediate}) to determine the posterior Beta distribution of the $j$-th compound in the $k$-th cluster.  Following Marlin (2008, Section 4.2), we introduce $z_{i,j}$ as
a missing data indicator,  assume missingness at random, and  then
recognize that the necessary counts record only data at non-missing entries:
\begin{eqnarray}
\label{Posterior6}
a_{k,j} = \alpha_0 + \sum_{i\in c_k} x_{i,j} \mathbbm{1}(z_{i,j} = 1), \quad b_{k,j} = \beta_0 + \sum_{i\in c_k}(1- x_{i,j}) \mathbbm{1}(z_{i,j} = 1).
\end{eqnarray}
Computations in Algorithms~\ref{Alg:greedy},~\ref{Alg:dpmm}, and~\ref{Alg:pel} proceed as usual with this
adjustment.

The GDSC data set is well structured to assess
the effect of missingness on IBR methods.  From the original 987 cell lines, 23 have complete data on all 304 drugs.  Data on the remaining 964 cell lines constitute our training data set $x_0$, which is sprinkled with missing data.   Each of the 23 remaining lines serves
as a novel target $i^*$ that each IBR method may operate on to prioritize bioactive compounds.   In revealing the
complete data on any $i^*$,  we have a test set from which 
NEF10 and ROCAUC metrics are derived; conveniently, the absence of missing data from the test set makes these metrics easier to compute. 

\begin{figure}[!ht]
\centering
\includegraphics[width=5.0in]{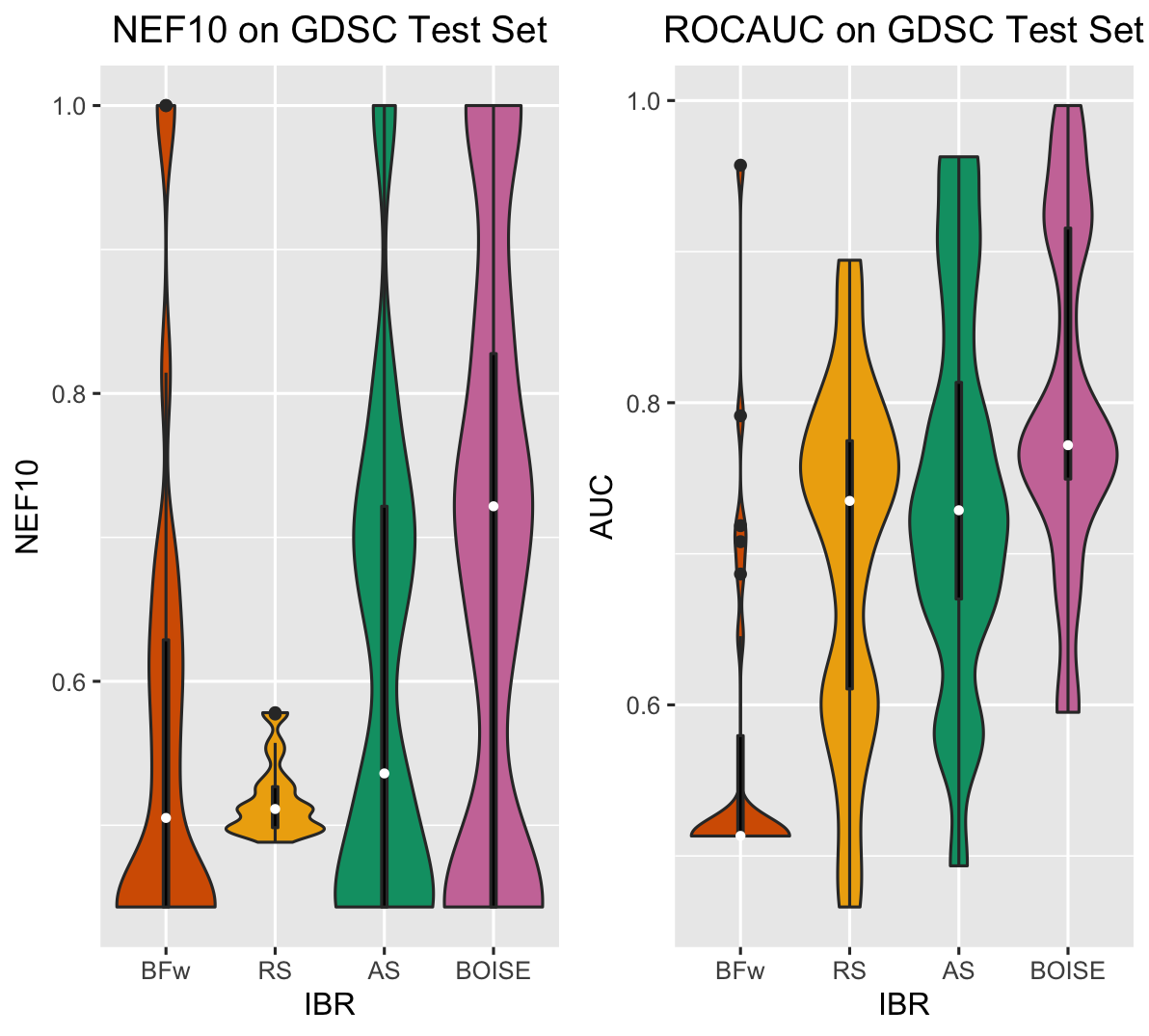}
\caption{\label{fig:mgdsc} 
{\bf Predictive performance of IBR methods in the presence of missing
data}.  Metrics and methods are as in Figures \ref{fig:pkis} and \ref{fig:gdsc}, but a single training set of 963 cell lines and 304 drugs from GDSC has 15.5\% missing entries.   For non-BOISE methods  (which require complete data), we performed matrix completion on the original data set  each method; BOISE, by contrast, operated on the available, incomplete
matrix. }
\end{figure}

Figure~\ref{fig:mgdsc} summarizes the predictive performance of BOISE, with
$n_A=8$, as well machine-learning IBRs adaptive selection (AS)  and regression selection (RS), and one of the baseline frequent-hitter rules.  The non-BOISE methods operate on completed data; we experimented with several matrix-completion tools and report  results from  Python package \verb+fancyimpute+ (\url{https://pypi.org/project/fancyimpute/}). Among the imputation methods provided by \verb+fancyimpute+, KNN imputation, with $K=3$, leads to the best retrospective results.  In spite of that selection for non-BOISE IBRs, BOISE 
is empirically stronger on both metrics.
The advantages are not statistically
significant by the  method used earlier, though the test-set size
is relatively small. For NEF10, for example,
 BOISE's  closest competitor has CI of $[-0.18, 0.036]$; for ROCAUC, the closest competitor has CI of $[-0.13, 0.016]$.
 More directly, on the 23 test-set targets, BOISE has
 better  NEF10 than the best competitor (AS) on
 7 of the targets, has worse numbers on 2, and gives the
 same top 10\% predictions as AS
 on the other 14. With
 ROCAUC, BOISE is better on 16 targets and worse on 7.

\section{Discussion}

Virtual screening trades
  biochemical experimentation for computer time.  It advances drug
  discovery efforts if the deployed algorithms  
  effectively encode  information on protein targets and drug compounds.
  Computational chemists are sometimes faced with the setting studied
  in this manuscript, wherein the target of interest 
  is known only to be a member of a class
  for which limited bioactivity data are available across a panel
  of drug compounds.  Effective though somewhat {\em ad hoc} machine-learning approaches have been developed for this experimental
  design problem.  For example, regression selection (RS) clusters
  initial targets via $K$-means clustering with $K$ obtained through cross
  validation.  Then it fits a regularized multinomial logistic regression
  to identify which compounds (the informers) best predict the cluster labels.
  The fact that RS and other machine-learning IBR methods perform better
  than domain-specific baseline methods suggests there is critical
  information in the bioactivity data available at the outset of the 
  experiment.      We reason that statistical approaches may 
  offer further insights, especially as we recognize the two-stage 
  problem structure and the opportunity for explicit risk minimization.
  
  The proposed BOISE IBR scheme shows strong predictive performance
  in  two retrospective empirical studies.  The source for 
  the improvements is not entirely clear. The 
  statistical model may be accurate, and then risk minimization
  does produce the most effective procedure.   It may be that model
  inaccuracies are less important than some key aspects of the computation,
  such as the fact that BOISE averages over uncertainties in 
  how targets should be clustered.   In any case the calculations
   reveal how Bayesian decision theory may operate in the realm 
   of virtual drug screening and what levels of prediction accuracy are
   possible.

Like for many Bayesian methods, a limitation of BOISE is its 
computational complexity.  Our prototype R code used approximately $20$ CPU hours on an Intel Core i5 processor to select an informer set with size $n_A=8$ for one  PKIS1 target, 
while $n_A=16$ required about $300$ CPU hours. For PEL$_1$ computation, the number of possible  intermediate data 
values $x_A$ increases as $2^{n_A}$.  The sampling strategy
in Algorithm~\ref{Alg:pel} avoids complete enumeration, but there is a trade-off between sample size and running time. For the PKIS1 retrospective analysis, 
we  used parallel computing available at the UW-Madison Center
for High Throughput Computing,
 completing calculations on 224 compute nodes in 2 weeks of wall time. 

Having developed a complete BOISE formulation, we can pursue
approximations that capture the essential structure with 
less computational effort.  The key step to compute PEL$_1$, for example,  is to evaluate $\mathbb E(\theta_{i^*j}\,|\, \mathcal C, x_0,x_A)$ in~(\ref{eq:thetatilde}), which consists of  the posterior probability $p_k = P(i^*\in c_k\,|\,\mathcal{C},x_0,x_A)$, and also the posterior expectation $E(\theta_{i^*j}\,|\, x_0,x_A,\mathcal{C},i^*\in c_k)$. 
Inspection shows that the informer set $A$ may have little impact on the second term, since for $j\notin A$,   intermediate data $x_A$ are not involved, while for $j\in A$ we already know the interaction of $j$ on $i^*$ through $x_A$. Hence the quantities $E(\theta_{i^*j}\,|\, x_0,x_A,\mathcal{C},i^*\in c_k)$ 
have a limited role in selecting top set $T^*(A,x_A,x_0)$, and  aspects of the distribution $p=(p_0,p_1,\cdots,p_K)$ alone may effectively score informer sets.  These probabilities
constitute the conditional distribution of the cluster label for target $i^*$ given $\mathcal C$, $x_A$ and $x_0$, and they
are relatively easy to compute. From Algorithm~\ref{Alg:pel} we see that $p_k$ can be calculated directly from $x_A$ and $x_0$ for each given $\mathcal C$, while $E(\theta_{i^*,j}\,|\, x_0, x_A)$ needs two rounds of averaging over all samples of $x_A$ and $\mathcal C$. Guided by the ID3 decision-tree method (Quinlan, 1986), we
take the entropy $H(p) = -\sum_{k=0}^K p_k \log_2 (p_k)$
and propose  $\mathcal{H}(A) = E\{ H(p) | x_0 \}$, 
which averages over $x_A$ and $\mathcal{C}$, as an objective
function to minimize in a simplified BOISE scheme.
The predictive performance of this entropy-based procedure is 
comparable to BOISE in the PKIS1 and GDSC retrospective studies,  but substantially better than other IBR methods (e.g., ROCAUC and NEF10 medians
in PKIS1 with $n_A=16$ were $0.951$ and $0.877$, respectively,  compared to values in Figure 2). However, the running time for the entropy-based method is dramatically reduced:   one  PKIS1 target takes $25$ CPU hours   compared to BOISE's  $300$ CPU hours. 

Making BOISE maximally applicable for broad biological targets may require addressing additional computational challenges.
In both retrospective analyses, the targets (protein kinases or cancer cell lines) were known in advance to be reasonably biologically similar.
However, in general a new target could be biologically distant from the targets with initial chemical screening data, for instance, a protein from a different family or an assay of a different cellular phenotype.
We can explore a wider variety of targets through retrospective analyses of bioassay data from PubChem (Kim {\em et al} 2019).
PubChem can also support testing BOISE's scalability to much larger datasets, as we can construct an initial bioactivity matrix containing partially-complete screening data for hundreds of targets and hundreds of thousands of compounds.
The PubChem-scale application may require algorithmic development to further improve the compute time and methodology development in order to degrade gracefully when the new target clusters poorly with initial targets.  

Some advantages may come from further consideration of loss functions and stochastic models.  For example, the current implementation allows overlap between $T$ and $A$, which might be handled differently for 
experiment prioritization. Also, the approach does not penalize molecules that have broad, non-specific activity.  It was not an issue in the examples presented here but a more elaborate parameter-space model
may be helpful  in larger cases.   Numerous factors warrant further study, and we hope the
   present framework is relevant 
   in this effort.

\section*{Acknowledgements}
This work was supported in part by National Institutes of Health awards R01GM135631, P50 DE026787, and P30CA14520-45, 
US  National Science Foundation grant 1740707, and the University of Wisconsin-Madison Office of the Vice Chancellor for Research and Graduate Education with funding from the Wisconsin Alumni Research Foundation.
The research was performed using the compute resources and assistance of the University of Wisconsin-Madison Biomedical Computing Group and the Center for High Throughput Computing. 

\section*{References}

\begin{list}{}{}



\item Berger, J.O. (1985) {\em Statistical decision theory and Bayesian analysis}. 2nd ed. Springer-Verlag, New York.

\item Berry, D.A. (2006) Bayesian Clinical Trials. {\em Nature Reviews Drug Discovery}, 5, 27-36.

\item Bajorath, J., Kearnes, S., Walters, W. P., Meanwell, N.A.,  Georg, G.I. and Wang, S. (2020)
Artificial Intelligence in Drug Discovery: Into the Great Wide Open. {\em Journal of Medicinal Chemistry}. doi.org/10.1021/acs.jmedchem.0c01077


\item Bretz, F., Hothorn, T. and Westfall, P. (2010) {\em Multiple Comparison Using R.} Chapman \& Hall, Boca Raton, FL.  


\item Drewry, D.H., Willson, T.M. and Zuercher, W.J. (2014) Seeding Collaborations to Advance Kinase Science with the GSK Published Kinase Inhibitor Set (PKIS). {\em Current Topics in Medicinal Chemistry}, 14(3), 340-342.






\item Henderson, N.C. (2015) {\em Methods for ranking and selection in large-scale inference.} Doctoral dissertation, University of Wisconsin-Madison, Madison, WI.

\item Hjort, N.L., Holmes, C., Muller, P. and Walker, S.G. (2010) {\em Bayesian 
 Nonparametrics}. Cambridge Series in Statistical and Probabilistic Mathematics.

\item Kim, S., Chen, J., Cheng, T., Gindulyte, A., He, J., He, S., Li, Q., Shoemaker, B.A., Thiessen, P.A., Yu, B., Zaslavsky, L., Zhang, J. and Bolton, E.E. (2019) PubChem 2019 update: improved access to chemical data. {\em Nucleic Acids Research}, vol. 47, issue D1, pages D1102–D1109.

\item Liu, J.S. (1994) The Collapsed Gibbs Sampler in Bayesian Computations with Applications to a Gene Regulation Problem. {\em Journal of the American Statistical Association}, 89:427, 958-966, DOI: 10.1080/01621459.1994.10476829


\item Liu, S., Alnammi, M., Ericksen, S.S., Voter, A.F., Ananiev, G.E., Keck, J.L., Hoffmann, F.M., Wildman, S.A. and Gitter, A. (2018) Practical model selection for prospective virtual screening. {\em Journal of Chemical Information and Modeling}, 59(1), 282-293.

\item MacEachern, S.N. (1994) Estimating normal means with conjugate style Dirichlet process prior. {\em Communications in Statistics - Simulation and Computation}, vol. 23, pp. 727-741.

\item Marlin, B.M. (2008) {\em Missing Data Problems in Machine Learning.} Doctoral dissertation, University of Toronto.


\item Neal, R.M. (2000) Markov Chain Sampling Methods for Dirichlet Process Mixture Models. {\em Journal of Computational and Graphical Statistics}, 9, 249-265.

\item Newton, M. A. and Geyer, C. J. (1994) Bootstrap recycling:  A Monte Carlo  algorithm  for  the  nested  bootstrap. {\em Journal of the American Statistical Association}, 89, 905-912. 

\item Parmigiani, G. and Inoue, L. (2009) {\em Decision theory: principles and approaches} (Vol. 812), West Sussex, England: John Wiley \& Sons.

\item Quinlan, J.R. (1986) Induction of Decision Trees. {\em Machine Learning}, 1, 81-106.



\item Robert, C. (2007) {\em The Bayesian choice: from decision-theoretic foundations to computational implementation.} Springer Science \& Business Media.

\item Sliwoski, G., Kothiwale, S., Meiler, J. and Lowe, E.W. (2014) Computational methods in drug discovery. {\em Pharmacological Reviews}, 66(1), pp.334-395.

\item Souza, P.C.T., Thallmair, S., Conflitti, P., Ram\'{i}rez-Palacios, C., Alessandri, R., Raniolo, S., Limongelli, V. and Marrink, S.J. (2020) Protein–ligand binding with the coarse-grained Martini model. {\em Nature Communications}, 11, 3714.

\item Trotter, H.F. and Tukey, J.W. (1954) Conditional Monte Carlo for normal samples. Symposium on Monte Carlo Methods.

\item Wald, A. (1950) {\em Statistical Decision Functions}. John Wiley \& Sons, Inc, New York.

\item Yang, W., Soares, J., Greninger, P., Edelman, E.J., Lightfoot, H., Forbes, S., Bindal, N., Beare, D., Smith, J.A., Thompson, I.R., Ramaswamy, S., Futreal, P.A., Haber, D.A., Stratton, M.R., Benes, C., McDermott, U. and Garnett, M.J. (2013) Genomics of Drug Sensitivity in Cancer (GDSC): a resource for therapeutic biomarker discovery in cancer cells. {\em Nucleic Acids Research}, vol. 41, pp. 955-961. 


\item  Zhang, H., Ericksen, S.S.,  Lee, C.,  Ananiev, G.E., Wlodarchak, N., Yu, P., Mitchell, J.C.,  Gitter, A.,  Wright, S.J., Hoffmann, F.M.,  Wildman, S.A. and 
Newton, M.A. (2019) Predicting kinase inhibitors using bioactivity matrix derived informer sets. {\em PLoS Computational Biology}, 15(8): e1006813.

\end{list}



\appendix 
\section{Proof of Theorem 1}

Considering the risk~(\ref{eq:risk}), it is sufficient to show that for any  fixed $x_A$, $x_0$, we have
\begin{eqnarray}
\label{eq:inequal1}
\int_\Theta  L\left( A, T \right) \, 
p(\theta |x_A,x_0) \, d\theta \geq \int_\Theta  L\left( A, T^* \right) \, 
p(\theta |x_A,x_0) \, d\theta.
\end{eqnarray} To confirm
this,  expand the loss function $L(A,T)$ and evaluate:
\begin{align*}
\int_\Theta  L\left( A, T \right) \, p(\theta |x_A,x_0) \, d\theta &= \int_\Theta  \left( n_T - \sum_{j\in T} \theta_{i^*j}\right) 
p(\theta |x_A,x_0) \, d\theta\\
&= n_T - \sum_{j\in T} \hat \theta_{i^*j}\\
&\geq n_T - \sum_{j\in T^*}  \hat \theta_{i^*j}\\
&= \int_\Theta  L\left( A, T^* \right) \, 
p(\theta |x_A,x_0) \,  d\theta.
\end{align*}

\section{Sampling clusterings given initial data}

A Dirichlet process mixture model (DPMM) has
data $x_i$ distributed $F(\cdot,\theta_i)$,
parameters $\{ \theta_i \}$ i.i.d. from $G$,
and uncertainty in $G$ following a Dirichlet
process with base measure $m_0 G_0$.
In BOISE, each $x_i$ is a multivariate random vector, and each component $x_{i,j}$ is an independent Bernoulli trial with success rate $\theta_{i,j}$. Therefore, $F(x_i,\theta_i)=
     \prod_{j=1}^n \theta_{i,j}^{x_{i,j}}\left(1-\theta_{i,j}\right)^{1-x_{i,j}}$
and  $G_0$ is a homogeneous Beta$(\alpha_0,\beta_0)$ on each component $\theta_{i,j}$.

Let $c(i)$  be the cluster label of  target $i$, $i = 1,2,\cdots,m$, and $\phi_k$ be the shared parameter in cluster $k$. Our goal is to sample $\mathcal C = \left(c(1), \cdots, c(m)\right)$ given $x_0$. Following the collapsing method from MacEachern (1994) and Neal (2000),
Algorithm~3, a Gibbs sampler uses:
\begin{equation*}
P(c(i) = k\,|\, c(-i), x_0) \propto \begin{cases}
		\frac{m_{-i,k}}{m-1+m_0}\int F(x_i,\phi_k)dH_{-i,k}(\phi_k), &\text{ if }
	    \exists \, c(j) = k, j\neq i\\
		\frac{m_0}{m-1+m_0}\int F(x_i,\phi)dG_0(\phi). & \text{ if }
		\nexists \, c(j)= k, j \neq i.
	\end{cases}
\end{equation*}
Here $m_{-i, k} = \sum_{j\neq i} \mathbbm{1}(c(j) = k)$ is defined to be the number of targets other than $i$ that are currently in cluster $k$, and $H_{-i,k}(\phi) = P\left(\phi_k\,|\,G_0,\{x_j:j\neq i, c(j)= k\} \right)$ is the posterior distribution of $\phi_k$ given $G_0$ and all other observations $x_j$ in cluster $k$ except $x_i$. We set $F(x_i, \phi)$ as above and $H_{-i, k}(\phi_k)$ as the product of independent Beta densities:
\begin{equation*}
    H_{-i, k}(\phi_k) = \prod_{j = 1}^n \frac{1}{B(a_{-i, k,j}, b_{-i,k,j})} \phi_{k,j}^{a_{-i,k,j}-1}(1-\phi_{k,j})^{b_{-i,k,j}-1}
\end{equation*}
where $a_{-i,k,j}$ and $b_{-i,k,j}$ are similarly defined as in~(\ref{eq:intermediate}). Notice that we use $k>0$ to denote an existing cluster and $k=0$ to denote a new cluster. The updating formula and corresponding algorithm for Gibbs sampler are:
\begin{equation}\label{eq:Postc}
P(c(i) = k| c(-i), x_0) \propto \begin{cases}
		\frac{m_{-i,k}}{m-1+m_0} \prod_{j=1}^n \left(\frac{a_{-i,k,j}}{a_{-i,k,j}+ b_{-i,k,j}}\right)^{x_{i,j}}\left(\frac{b_{-i,k,j}}{a_{-i,k,j}+ b_{-i,k,j}}\right)^{1-x_{i,j}} &\text{if } k > 0
	    \\
		\frac{m_0}{m-1+m_0}\prod_{j=1}^n \left(\frac{\alpha_0}{\alpha_0 + \beta_0}\right)^{x_{i,j}}\left(\frac{\beta_0}{\alpha_0+\beta_0}\right)^{1-x_{i,j}}. & \text{if }
		k = 0.
	\end{cases}
\end{equation}




\begin{algorithm}
\caption{DPMM clustering}\label{Alg:dpmm}
\begin{algorithmic}[1]
\State Set the prior mass $m_0$ and  hyperparameters
$\alpha_0$ and $\beta_0$
\State Initialize $c(1),\cdots,c(m)$ ;
\State Set sample size of clustering assignments $M$, and gaps between successive draws $N$
\While{Sample size $< M$}
    \For{$i=1, \cdots, m$}
        \State Update $i$th cluster label from  $P(c(i)\,|\,c(-i),x_0)$ in~(\ref{eq:Postc})
    \EndFor
\State Record  clustering assignment after every $N$ 
cycles.
\EndWhile
\end{algorithmic}
\end{algorithm}

We follow the empirical Bayes principle to choose hyperparameters. We select $\alpha_0 = {\rm mean}(x_0)$ and $\beta_0 = 1 - \alpha_0$ and select prior mass $m_0$ to make prior cluster numbers as close to posterior cluster numbers as possible. For PKIS1 data, we select $m_0 = 15$ and $\alpha_0 = 0.066$; For GDSC complete data, we select $m_0 = 3$ and $\alpha_0 = 0.067$. The sample size $M$ is $100$ and the thinning step $N$ is $50$ for both data sets,
which shows adequate mixing in MCMC output analysis. 


\section{Recycling algorithm}
To compute the optimal top set $T^*$ associated with hypothetical
intermediate data $x_A$, we need to compute posterior means
$\hat \theta_{i^*,j}$ for all compounds $j$.
By tilting the expectation as in importance sampling,  we have, summing over all partitions
$\mathcal{C}$ of initial proteins $I$,
\begin{eqnarray*}
\hat \theta_{i^*,j} &=& E\left( \theta_{i^*,j} | x_0, x_A \right) \\
&=&\sum_{\mathcal{C}} \left( \tilde \theta_{i^*,j} \right) \, p( \mathcal{C} | x_0, x_A ) \\ 
 &= & \sum_{\mathcal{C}} \left( \tilde \theta_{i^*,j} \right) \frac{ p( \mathcal{C} | x_0, x_A ) }{ p( \mathcal{C} | x_0 ) }
  \, p( \mathcal{C} | x_0 )   \\
  &=& \sum_{ \mathcal{C}} \left( \tilde \theta_{i^*,j} \frac{ p( x_A |x_0, \mathcal{C} ) }{ p( x_A | x_0) } \right)
  \, p( \mathcal{C} | x_0 ). 
\end{eqnarray*}
where, because of~(\ref{eq:intermediate}),
\begin{eqnarray}
\label{eq:pxagx0C}
p(x_A|x_0,\mathcal C) = \sum_{k=0}^K\frac{m_k}{m+m_k} \prod_{j\in A} \left(\frac{a_{k,j}}{a_{k,j}+b_{k,j}}\right)^{x_{i^*,j}}\left(\frac{b_{k,j}}{a_{k,j}+b_{k,j}}\right)^{1-x_{i^*,j}}.
\end{eqnarray}
Thus,   quantities $\hat \theta_{i^*,j}$
are also all expectations of modified objects with respect to the original posterior $p(\mathcal{C}|x_0)$, and so we may re-use the Monte Carlo samples to approximate for each predictive sample $x_A$.
The only trick is to get $p(x_A|x_0)$, which we can get directly by averaging
the values $p(x_A|x_0,\mathcal{C})$  
over these sampled clusterings.  If $\mathbb C$ is a collection of 
$N$ clusterings  $\mathcal{C}$'s sampled from $p(\mathcal{C}|x_0)$, then we approximate 
$\hat \theta_{i^*,j}$ by
\begin{eqnarray}
\label{eq:recycle}
\frac{1}{N} \sum_{\mathcal{C}\in \mathbb{C}} \left\{\sum_{k=0}^K p_k \left( \frac{a_{k,j}+x_{i^*,j}\mathbbm{1}\left(j\in A\right)}{a_{k,j}+b_{k,j}+\mathbbm{1}\left(j\in A\right)} \right)\right\} \left( \frac{p(x_A|x_0,\mathcal C)}{\frac{1}{N}\sum_{\mathcal{C}' \in \mathbb{C}} p(x_A|x_0,\mathcal{C}')}  \right)
\end{eqnarray}
Our use of recycled samples is similar
to their use other contexts  (e.g., Newton and Geyer 1994;
Trotter and Tukey, 1954).  Pseudocode is in Algorithm~\ref{Alg:pel}.

\begin{algorithm}
\caption{PEL Computation}\label{Alg:pel}
\hspace*{\algorithmicindent} \textbf{Input:} Initial data $x_0$, informer set $A$, size of top set $n_T$,  samples $\mathbb C$ from $p(\mathcal{C}\,|\,x_0)$.\\
\hspace*{\algorithmicindent} \textbf{Output:} ${\rm PEL}_1(x_0,A)$ as score of informer set $A$. 
\begin{algorithmic}[1]
\For{each $\mathcal{C}$ in $\mathbb C$}
        \State Sample $\mathbb X_{A,\mathcal C}$ from $p(x_A|\mathcal{C},x_0)$ as~(\ref{eq:intermediate}) \Comment{Intermediate data sample}
\EndFor
\State Set $\mathbb X_A = \bigcup \mathbb X_{A, \mathcal{C}}$
\For{each distinct $x_A$ in $\mathbb X_{A}$}
    \For{each $\mathcal C$ in $\mathbb C$}
        \State Calculate $\tilde \theta_{i^*,j} = \mathbb E(\theta_{i^*j}|\mathcal C, x_0,x_A)$ using~(\ref{eq:thetatilde});
        \State Calculate $p(x_A\,|\,x_0, \mathcal C)$ using~(\ref{eq:pxagx0C}).
	\EndFor
	\State Calculate $\hat \theta_{i^*,j}=E( \theta_{i^*,j} | x_A, x_0)$ for each $j$
	   using~(\ref{eq:recycle})
    \State Calculate PEL$_2(x_0,A,x_A)$ loss as in~(\ref{eq:pel2})
\EndFor
    \State Calculate PEL$_1(x_0,A)$ by averaging PEL$_2(x_0,A,x_A)$ over all $x_A \in \mathbb X_A$.
\end{algorithmic}
\end{algorithm}

\section{Normalized enrichment factor (NEF10)}

Enrichment factor (EF) is a commonly used metric in context of virtual screening. It reflects how much increase in active compounds compared to random selection. EF is actually a scaled form of TPR: After IBR ranking, we can relabel compounds $j=1,2, \cdots, n$
by highest to lowest priority for further testing on target $i^*$.  The
10\% enrichment 
factor is
\begin{eqnarray}
\label{eq:EF}
{\rm EF10}_{i^*} = \frac{\sum_{j=1}^{ \lfloor n/10 \rfloor } x_{i^*,j}}{\lfloor n/10 \rfloor}/\frac{\sum_{j=1}^n x_{i^*,j}}{n}
\end{eqnarray}
In~(\ref{eq:EF}), we can see that EF is influenced by the number of active compounds of the target. Therefore a normalized EF (NEF) is introduced in Zhang {\em et al.} (2019) to make better comparison across targets with different active ratios:
\begin{eqnarray}
\label{eq:NEF}
{\rm NEF10}_{i^*} = \left(1+\frac{{\rm EF10}_{i^*} - {\rm EFbase}}{{\rm EF10}_{max} -{\rm EFbase} }\right)/2
\end{eqnarray}
where ${\rm EFbase} = 1$ corresponds to random guessing, and ${\rm EF10}_{max}$ is the maximum theoretical value of ${\rm EF10}$. The ${\rm NEF10}$ value is between $0$ and $1$ with random guessing at $0.5$.

\end{document}